\documentclass[epj]{svjour}

\usepackage{graphics}
\usepackage{amssymb}
\usepackage{amsmath}
\usepackage{overpic}

\newcommand{\abs}[1]{\ensuremath{\left\vert#1\right\vert}}
\newcommand{\ket}[1]{\ensuremath{|#1\rangle}}

\title{Generalized Inverse Participation Numbers in Metallic-Mean Quasiperiodic Systems}
\titlerunning{Generalized Inverse Participation Numbers in Metallic-Mean Systems}
\author{Stefanie Thiem\thanks{\email{stefanie.thiem@physik.tu-chemnitz.de}} and Michael Schreiber}
\institute{Institut f\"ur Physik, Technische Universit\"at Chemnitz, D-09107 Chemnitz, Germany}

\abstract{
From the quantum mechanical point of view, the electronic characteristics of quasicrystals are determined by the nature of their eigenstates. A practicable way to obtain information about the properties of these wave functions is studying the scaling behavior of the generalized inverse participation numbers $Z_q \sim N^{-D_q(q-1)}$ with the system size $N$. In particular, we investigate $d$-dimensional quasiperiodic models based on different metallic-mean quasiperiodic sequences. We obtain the eigenstates of the one-dimensional metallic-mean chains by numerical calculations for a tight-binding model. Higher dimensional solutions of the associated generalized labyrinth tiling are then constructed by a product approach from the one-dimensional solutions. Numerical results suggest that the relation $D_q^{d\mathrm{d}} = d D_q^\mathrm{1d}$ holds for these models. Using the product structure of the labyrinth tiling we prove that this relation is always satisfied for the silver-mean model and that the scaling exponents approach this relation for large system sizes also for the other metallic-mean systems.
}

\begin{document}

\maketitle

\section{Introduction}

Since the discovery of quasicrystals \cite{PhysRevLett.1984.Shechtman,PhysRevLett.1985.Ishimasa,PhysRevLett.1987.Wang} various experimental investigations have revealed the rather uncommon electrical, magnetic and optical characteristics of real quasicrystals \cite{UsefulQuasicrystals,PhysicalProperties.1999.Stadnik} and motivated extensive research to obtain a better theoretical understanding of their structure and properties. Today many exact results are known for one-dimensional quasicrystals \cite{MathQuasi.2000.Damanik}, but the characteristics of wave functions in two or three dimensions are understood to much lesser degree. Also numerical investigations often deal only with finite, relatively small systems or periodic approximants, which allows only the estimation of a few properties of ideal quasicrystals \cite{Quasicrystals.2003.Grimm,PhysRevB.1992.Passaro,ICQuas.1998.Rieth}.

An alternative approach is based on the study of $d$-dimensional models with separable Hamiltonians such that higher dimensional, quantum mechanical solutions can be directly derived from the one-dimensional case, allowing the numerical consideration of large systems up to $10^8$ sites. The best studied model within this approach is based on the Fibonacci sequence \cite{Ferro.2004.Ilan,JAC.2002.Lifshitz,JPhysFrance.1989.Sire}, whereas in this paper we concentrate on the more general class of metallic-mean quasiperiodic sequences \cite{JPhysFrance.1989.Sire2,NonLinAnal.1999.Spinadel,PhysRevB.2000.Yuan}, which has been less investigated so far. The sequences describe the weak and strong couplings of atoms in a quasiperiodic chain and the one-dimensional eigenstates are obtained by numerical diagonalization of the respective tight-binding Hamiltonian. Higher-dimensional quasiperiodic tilings, the so-called generalized labyrinth tilings,  are constructed as a direct product of these chains and their eigenenergies and  eigenstates can be directly calculated by multiplying the energies $E$ and wave functions $\Psi$ of the quasiperiodic chain, respectively \cite{JPhysCS.2010.Thiem}.

While for an electron in a one-dimensional quasiperiodic system many examples lead to purely singular continuous energy spectra and multifractal eigenstates \cite{PhysRevB.1987.Kohmoto,JStatPhys.1989.Suto,CMathPhys.1989.Bellissard}, in the two-dimensional labyrinth tiling the Lebesgue measure of the energy spectrum shows a transition from zero to a finite value with increasing coupling parameter $v$ \cite{EurophysLett.1989.Sire}. Regardless of this transition of the energy spectrum, the numerical results show that the wave functions remain multifractals in two and three dimensions and do not become extended with increasing values of $v$. To address this, we investigate the scaling behavior of the generalized inverse participation numbers $Z_q$ of the eigenstates and their dependency on the dimension $d$ in more detail. Further, we mathematically prove that the generalized dimensions associated to the inverse participation numbers $Z_q$ satisfy the relation $D_q^{d\mathrm{d}} = d D_q^\mathrm{1d}$ for the silver-mean model and asymptotically approach this relation for the other metallic-mean systems.

This paper is organized as follows: In Sec.~\ref{sec:construction} we introduce the inflation rule for the quasiperiodic metallic-mean chains and discuss the properties of the corresponding one-dimensional eigenstates as well as the construction of the eigenstates of the associated higher-dimensional labyrinth tilings. Section \ref{sec:partnumbers} then focuses on the inverse participation numbers by discussing numerical results and proving the above mentioned relation between the generalized scaling exponents $D_q$ in different dimensions. This is followed by a brief summary of our results.

\section{Eigenstates and Wave Functions of Generalized Labyrinth Tilings}\label{sec:construction}

The construction of the labyrinth tiling is based on the so-called metallic-mean quasiperiodic sequences, which for a parameter $b$ are defined by the inflation rule
\begin{equation}
 \label{equ:octonacci.1} \mathcal{P} =
 \begin{cases}
   w \longrightarrow s \\
   s \longrightarrow sws^{b-1}
 \end{cases}
\end{equation}
starting with the symbol $w$. After $a$ iterations we obtain the $a$th order approximant $\mathcal{C}_a$ of the quasiperiodic chain. The length $f_a$ of an approximant $\mathcal{C}_a$ is also given by the recursive rule $f_a = b f_{a-1} + f_{a-2}$ with $f_0 = f_1 = 1$. Thereby, the ratio of the lengths of two successive iterants in the limit $ \lim_{a \rightarrow \infty} f_a / f_{a-1} = \lambda$ equals different metallic means depending on the parameter $b$ with a continued fraction representation $\lambda = [\bar{b}] = [b,b,b,...]$. This leads to the well known Fibonacci sequence for $b=1$ with the golden mean $\lambda_{\mathrm{Au}} = [\bar{1}]=(1+\sqrt{5})/2$, while $b=2$ results in the octonacci sequence with silver mean $\lambda_{\mathrm{Ag}} = [\bar{2}] = 1+\sqrt{2}$ and $b=3$ corresponds to the bronze mean $\lambda_{\mathrm{Bz}} = [\bar{3}] = (3+\sqrt{13})/2$ \cite{JPhysA.1989.Gumbs}.

\subsection{Eigenstates of metallic-mean chains}\label{sec:1dstates}

Solving the time-independent Schr\"odinger equation for the quasiperiodic systems with zero on-site potentials
\begin{equation}
 \label{equ:octonacci.8}
 \mathcal{H} \ket{\Psi^i} = E^i \ket{\Psi^i} \Longrightarrow E^i \Psi_l^i = t_{l-1,l} \Psi_{l-1}^i + t_{l,l+1} \Psi_{l+1}^{i} \;,
 \end{equation}
we obtain the discrete energy values $E^i$ and the wave functions $\ket{\Psi^i} = \sum_{l=1}^{f_a+1} \Psi_l^i \ket{l}$ represented in the orthogonal basis states $\ket{l}$ associated to a vertex $l$. The hopping parameter $t$ in the Schr\"odinger equation is given according to the quasiperiodic sequence $\mathcal{C}_a$ with $t_{s} = 1$ for a strong bond and $t_{w} = v$ for a weak bond ($0 \le v \le 1$) \cite{Quasicrystals.2003.Grimm,PhysRevB.2000.Yuan}. Applying free boundary conditions the number of vertices is $N_a = f_a + 1$.

The results show that the eigenvalues are symmetric with respect to $0$. Thus, for even system sizes $N_a$ all energy values $E$ have a symmetric value $-E$, but for odd $N_a$ there is one state $E^M = 0$, which has no corresponding state. Here $M = \lfloor N_a/2 \rfloor + 1$, assuming that the eigenstates are labeled according to increasing eigenenergies.

Additionally, the eigenfunctions possess a symmetry. The eigenstate $\Psi$ with the eigenvalue $E$ and the eigenstate $\widetilde{\Psi}$ with the corresponding eigenvalue $-E$ only differ by an alternating sign depending on the vertex $l$ according to
 \begin{equation}
  \label{equ:octonacci.6}
  \widetilde{\Psi}_l = (-1)^l \Psi_l \;.
 \end{equation}
Further, for odd $N_a = 2M-1$ the eigenvector $\Psi_l^M$ associated to the eigenvalue $E^M = 0$ has a special structure, namely $\Psi_l^M$ vanishes either on all odd or on all even sites $l$ (cp.\ \cite{PhysRevB.2000.Yuan}), i.e.\
 \begin{equation}
  \label{equ:octonacci.7}
   \Psi_{l\bmod 2 \equiv 0}^{M^-} = 0 \qquad\quad  \textrm{and} \quad\qquad \Psi_{l\bmod 2 \equiv 1}^{M^+} = 0 \;.
 \end{equation}

\subsection{Eigenstates of the generalized labyrinth tilings}\label{sec:labyrinth}

The generalized labyrinth tilings in $d$ dimensions are constructed from $d$ quasiperiodic chains $\mathcal{C}_a$ perpendicular to each other, where the bonds are given by the diagonals. Hence, depending on the starting point the grid decomposes into $2^{d-1}$ separate grids. Each of these grids corresponds to a finite $a$th order approximant $\mathcal{L}_a$ of the generalized labyrinth tiling $\mathcal{L}$ \cite{JPhysFrance.1989.Sire2,PhysRevB.2000.Yuan}. In the case of the octonacci sequence all $2^{d-1}$ grids are identical. For other inflation rules the different approximants are slightly shifted against each other and mainly differ at the boundaries. These similarities originate from hidden mirror symmetries in the quasiperiodic chains. While the octonacci chain is perfectly mirror symmetric, for the other chains mirror symmetry can be achieved by neglecting a few symbols at one end of the sequence (e.g. for the Fibonacci chain the last two symbols on the right) and by interchanging two consecutive bonds for the case $b > 2$ (i.e.\ a phason flip) \cite{JPhysCS.2010.Thiem}.

The eigenstates of the generalized labyrinth tiling in $d$ dimensions are then constructed as the product of the eigenstates of $d$ one-dimensional chains. For instance in two dimensions we use $E^{ij} = E^{1i} E^{2j}$ and  $\Phi_{lm}^{ij} \sim \Psi_{l}^{1i} \Psi_{m}^{2j}$ \cite{PhysRevB.2000.Yuan}. The indices $i$ and $j$ enumerate the eigenvalues $E$ in ascending order and $l$ and $m$ represent the coordinates of the vertices. However, due to the symmetries of the eigenfunctions of Eq.\ \eqref{equ:octonacci.6}, some of the wave functions $\Phi_{lm}^{ij}$ and the related eigenvalues $E^{ij}$ are identical and then only one of them is allowed to be considered.

In two dimensions for even chain lengths $N_a$ the system size is $N_a^2/2$ and a valid combination of eigenstates is
 \begin{equation}
  \label{equ:labyrinth.4a}
  \left\{ E^{ij} \,|\, 1\le i \le N_a/2 \wedge 1 \le j \le N_a \right\}\;.
 \end{equation}
The normalization of the wave functions results in
 \begin{equation}
  \label{equ:labyrinth.3a}
  \Phi_{lm}^{ij} = \sqrt{2}  \Psi_{l}^{i}  \Psi_{m}^{j} \;.
 \end{equation}
For odd chain length $N_a$ the selection of the eigenstates is more complicated due to the special structure of the eigenfunction for $E^M = 0$. This results in the set of two-dimensional energy values
 \begin{multline}
  \label{equ:labyrinth.4b}
  \{ E^{ij} \,|\, \left(1\le i< M \wedge 1 \le j \le N_a \right) \vee \\ \left( i=M \wedge 1 \le j \le M \right) \} \;.
 \end{multline}
Further, the normalization of the wave functions is given by
 \begin{equation}
  \label{equ:labyrinth.3b}
  \Phi_{lm}^{ij} =
  \begin{cases}
     \Psi_{l}^{i}  \Psi_{m}^{j}  \qquad\quad\; i=j=M\\
     \sqrt{2} \Psi_{l}^{i} \Psi_{m}^{j}  \qquad \textrm{otherwise}
  \end{cases} \qquad .
 \end{equation}
The special treatment of the case $i=j=M$ is due to the fact that the wave function $\Phi_{lm}^{MM}$ lives only on one of the two different labyrinth tilings, $\mathcal{L}$ or $\mathcal{L}^\star$.

We analogously compute the eigenenergies and wave functions in higher dimensions. For even $N_a$ the eigenvalues $E^{i\ldots j k} = E^i \ldots E^j E^k$ of the generalized labyrinth tiling $\mathcal{L}_a^{d\mathrm{d}}$ in $d$ dimensions are described by the set
 \begin{multline}
  \label{equ:labyrinth3D.3a}
   \{ E^{i\dots j k} \,|\, 1\le i \le N_a/2 \wedge \ldots \wedge 1 \le j \le N_a/2 \wedge  \\ 1 \le k \le N_a \} \;.
 \end{multline}
This equation results from the symmetry of the wave functions given in Eq.\ \eqref{equ:octonacci.6}. Due to the special structure of the eigenfunctions with the eigenvalue $E^M = 0$, for odd $N_a$ the allowed set of eigenstates is much more complex. For instance, in three dimensions a valid combination is given by
  \begin{align}
    \label{equ:labyrinth3D.3b}
   \{ E^{ijk} \,|\, & \left(1\le i < M \wedge 1 \le j < M \wedge 1\le k \le N_a \right) \vee \nonumber\\
                    &  \left(i = M \wedge 1 \le j \le M \wedge 1 \le k \le M \right)  \vee \\
                    & \left(1 \le i < M \wedge j = M \wedge 1 \le k \le M \right)  \} \;. \nonumber
  \end{align}
Further, the normalization constants of the wave functions $\Phi_{lmn}^{ijk}$ for even system sizes $N_a$ are
 \begin{equation}
  \label{equ:labyrinth3D.4a}
  \Phi_{lmn}^{ijk} = \sqrt{2^{d-1}} \Psi_{l}^{i}  \Psi_{m}^{j}  \Psi_{n}^{k}
 \end{equation}
in analogy to two dimensions. For odd $N_a$ and three dimensions we obtain
 \begin{equation}
  \Phi_{lmn}^{ijk} =
   \begin{cases}
     \Psi_{l}^{i}  \Psi_{m}^{j}  \Psi_{n}^{k}  &  i=j=k=M\\
     \sqrt{2}  \Psi_{l}^{i}  \Psi_{m}^{j} \Psi_{n}^{k} & \textrm{two indices } i, j, k \textrm{ equal } M\\
     2  \Psi_{l}^{i}  \Psi_{m}^{j}  \Psi_{n}^{k} &    \textrm{otherwise} \;.
   \end{cases}  \label{equ:labyrinth3D.4b}
 \end{equation}

For larger approximants the wave functions approach each other because the influence of the boundaries and of the phason flip for $b > 2$ vanishes \cite{JPhysCS.2010.Thiem}. This characteristic is useful for analytical considerations of properties of the eigenstates of the general labyrinth tiling.

\section{Scaling Behavior of Generalized Inverse Participation Numbers}\label{sec:partnumbers}

\begin{figure*}
 \centering
    \begin{overpic}[height=6.3cm]{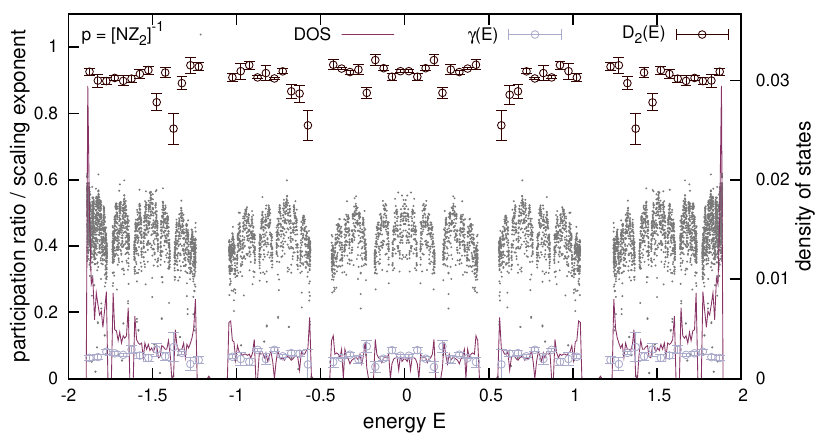} \put(2,2){\footnotesize{(a)}} \end{overpic}
    \begin{overpic}[height=6.3cm]{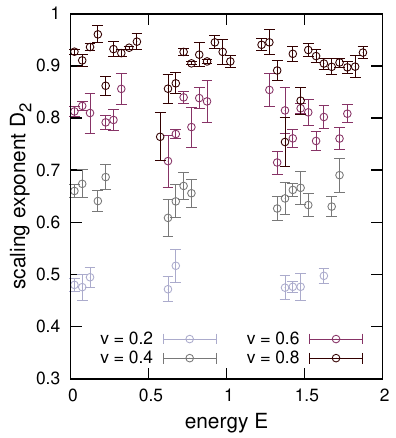} \put(2,3){\footnotesize{(b)}} \end{overpic}
  \caption{Energy dependence of the scaling exponent $D_2$ of the inverse participation number $Z_2(E)$ averaged over an energy interval $[E-\Delta E/2, E+\Delta E/2]$ with $\Delta E = 0.05$ for the silver-mean chain $\mathcal{C}_{11}^{\mathrm{Ag}}$ for $v = 0.8$ (a) and for different values of $v$ (b). In (a) also the values of the participation ratio $p(\Psi)$ and the scaling exponent $\gamma(E)$ of the averaged participation ratio $p(E)$ as well as the density of states (DOS) are shown.}
  \label{fig:gammaE}
\end{figure*}

In a $d$-dimensional system with discrete positions $\textbf{r}$ the generalized inverse participation number of an eigenstate $\Phi_{\textbf{r}}$ is defined for a parameter $q$ as
  \begin{equation}
  \label{equ:participation.1}
  Z_q(\Phi) = \sum_{\textbf{r}} \abs{\Phi_{\textbf{r}}}^{2q}\;.
 \end{equation}
This is associated to the scaling of the $q$th moment of the eigenstates $\Phi_\textbf{r}$ according to a power law
 \begin{equation}
  Z_q(\Phi) \sim N^{-D_q(q-1)}
 \end{equation}
with the linear dimension of the system $N$ and the generalized dimensions $D_q$.

In particular, the inverse of $Z_2$ equals the total number of sites for which the probability measure of the wave function given by $\abs{\Phi_{\textbf{r}}}^{2}$ is significantly different from zero and hence is a measure for the localization of the wave function $\Phi_{\textbf{r}}$. Investigating the scaling behavior of the participation ratio
 \begin{equation}
 p(\Phi) = \frac{Z_2^{-1}(\Phi)}{V} \sim N^{-d\gamma}
  \end{equation}
in the volume $V$, the wave function of a localized state is characterized by the scaling exponent $\gamma = 1$, and $\gamma = 0$ corresponds to an extended state. Intermediate values of $\gamma$ ($0 < \gamma < 1$) indicate fractal eigenstates, which are neither extended over the whole system nor completely localized at a certain site and show self-similar patterns \cite{PhysRevB.1987.Kohmoto,PhysRevB.1998.Repetowicz,Thesis.2008.Thiem}.

Within a small energy interval the participation ratios usually fluctuate over a certain range and the average participation ratio is often also energy dependent. For instance, for octagonal tiling models, for the Penrose tiling and for the two- and three-dimensional Rauzy tilings the participation ratios are found to be smaller at the center of the energy spectrum than at its edges, while for the three-dimensional Ammann-Kramer-Neri tiling the inverse behavior can be observed \cite{Quasicrystals.2003.Grimm,PhysRevB.1992.Passaro,PhilMag.2007.Jagannathan,PhysRevB.2002.Triozon}. However, to get information about the spatial distribution of the wave functions for macroscopic systems one has to compute the scaling behavior of the participation ratios with the system size. In particular, numerical results for the Penrose tiling and the Ammann-Kramer-Neri tiling revealed that the scaling exponents $\gamma(E)$ show a qualitatively similar dependence on the energy as the participation ratios \cite{Quasicrystals.2003.Grimm}.

\begin{figure*}
 \centering
    \begin{overpic}[width=0.32\textwidth]{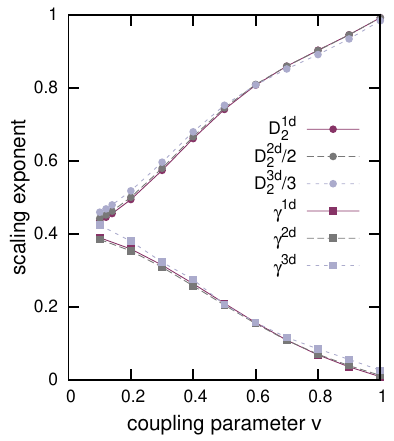} \put(2,3){\footnotesize{(a)}} \end{overpic}
    \begin{overpic}[width=0.32\textwidth]{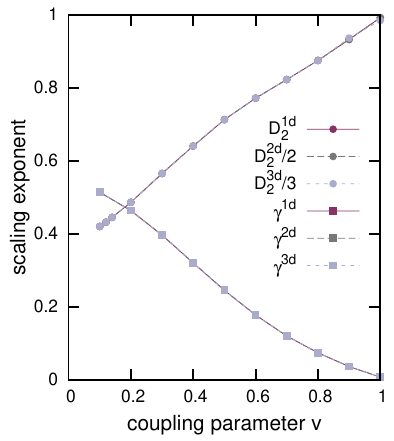} \put(2,3){\footnotesize{(b)}} \end{overpic}
    \begin{overpic}[width=0.32\textwidth]{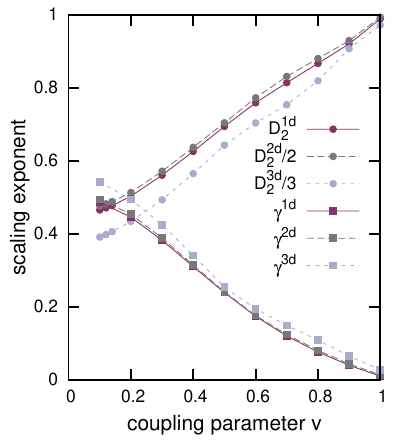} \put(2,3){\footnotesize{(c)}} \end{overpic}
  \caption{Scaling exponents $\gamma$ of the participation ratio $\langle p\rangle$ and $D_2$ of $\langle Z_2 \rangle$ averaged over the complete energy spectrum in one, two and three dimensions for the golden-mean model (a), the silver-mean model (b), and the bronze-mean model (c).}
  \label{fig:pscaling_exp}
\end{figure*}

In contrast to the former discussed systems, for the metallic-mean models the participation ratios and inverse participation numbers do not show any significant trend over the whole energy spectrum as visualized in Fig.\ \ref{fig:gammaE}. This might be related to the same number of nearest neighbors for every vertex in these tilings. In order to compare the characteristics of the eigenstates for different dimensions we consider the inverse participation numbers $\langle Z_2 \rangle$ and the participation ratio $\langle p \rangle$ averaged over the complete energy spectrum. Exemplarily, we show numerical results for the dependency of the corresponding scaling exponents $D_2$ and $\gamma$ on the coupling constant $v$ for the golden-, silver-, and bronze-mean model in one, two and three dimensions in Fig.\ \ref{fig:pscaling_exp}. This indicates that both scaling exponents are independent of the dimension \cite{JPhysCS.2010.Thiem}. The deviations in three dimensions are most likely caused by the consideration of relatively small sizes of the associated one-dimensional chains with $N_{12}^\mathrm{Au} = 234$ and $N_5^\mathrm{Bz} = 143$ sites. Similar results can also be found for the scaling exponents $D_q$ of the average generalized inverse participation numbers $\langle Z_q \rangle$.

In the following section we show that for the octonacci sequence and its associated labyrinth tilings
the scaling exponents $D_q^{d\mathrm{d}}$ of the average generalized inverse participation numbers $\langle Z_q \rangle$ in $d$ dimensions are multiples of the one-dimensional scaling exponents $D_q^{1\mathrm{d}}$. Further, we show that for the other quasiperiodic systems this relation is approached for large system sizes $N_a$. All following derivations can be equivalently applied to the scaling of the average participation ratio $\langle p \rangle$, which is just a function of $Z_2$, and hence it can be derived that the scaling exponent $\gamma$ is independent of the dimension.

\subsection{Special case -- silver-mean model}

We start with the silver-mean model, for which all sequences are palindroms and only even system sizes occur. Thus, the lattice resolves into $2^{d-1}$ identical lattices in $d$ dimensions \cite{JPhysCS.2010.Thiem}. This allows us to decompose the calculations of the higher-dimensional inverse participation numbers $Z_q(\Phi)$ for a wave function $\Phi$ with arbitrary dimension into a product of inverse participation numbers $Z_q(\Psi)$ of the one-dimensional wave functions $\Psi$ \cite{PhysRevB.2005.Cerovski} according to:
\begin{subequations}
 \label{equ:participation}
 \begin{align}
  Z_q(&\Phi^s) = \sum_{\textbf{r} \in \mathcal{L}} \abs{\Phi_\textbf{r}^s}^{2q} \label{equ:participation.2a}\\
             & = \sum_{l,m,\ldots,n \in \mathcal{L}} \abs{\sqrt{2^{d-1}}\Psi_l^i \Psi_m^j \ldots \Psi_n^k}^{2q} \label{equ:participation.2b}\\
             & \stackrel{\eqref{equ:octonacci.6}}{=} \frac{2^{q(d-1)}}{2^{d-1}} \sum_{l=1}^{N_a} \abs{\Psi_l^i}^{2q} \sum_{m=1}^{N_a} \abs{\Psi_m^j}^{2q} \ldots \sum_{n=1}^{N_a}\abs{\Psi_n^k}^{2q} \label{equ:participation.2c}\\
             & = 2^{(d-1)(q-1)} Z_q(\Psi^i) Z_q(\Psi^j) \ldots Z_q(\Psi^k) \label{equ:participation.2d}\;.
 \end{align}
 \label{equ:participation.2}
\end{subequations}
The step from Eq.\ \eqref{equ:participation.2b} to Eq.\ \eqref{equ:participation.2c} uses the fact that the grid is $2^{d-1}$-partite in $d$ dimensions, i.e.\ it resolves into $2^{d-1}$ identical labyrinth tilings, which only differ by their starting point for the construction rule. Consequently, each of these tilings contributes with the same amount to the inverse participation number. In Eq.\ \eqref{equ:participation.2b} we only sum over the vertices of one of these tilings. Due to this relation the summation in Eq.\ \eqref{equ:participation.2c} can be separated into $2^{d-1}$ parts, which each yield the same result, and hence, we obtain the desired product structure.

Assuming that the average inverse participation number obeys the scaling law $\langle Z_q^{1\mathrm{d}} \rangle \sim N^{-\tau_q^{1\mathrm{d}}}$ with $\tau_q^{1\mathrm{d}} = D_q^{1\mathrm{d}}(q-1)$ for the one-dimensional octonacci chain, we can derive scaling expressions for the associated $d$-dimensional labyrinth tiling starting with the general definition of the generalized inverse participation numbers:

{\allowdisplaybreaks
\begin{subequations}
 \begin{align}
  \langle & Z_q(\Phi^s) \rangle  =  \frac{1}{ V_a^{d\mathrm{d}}} \sum_{s \in \mathcal{L}} Z_q(\Phi^s) \label{equ:proof1a}\\
  & \stackrel{\eqref{equ:participation}}{=}  \frac{2^{q(d-1)}}{N_a^d} \sum_{i=1}^{N_a/2} Z_q(\Psi^i) \ldots \sum_{j=1}^{N_a/2} Z_q(\Psi^j) \sum_{k=1}^{N_a}  Z_q(\Psi^k) \label{equ:proof1b}\\
  & \stackrel{\eqref{equ:octonacci.6}}{=}  \frac{2^{q(d-1)}}{N_a^d} \frac{1}{2^{d-1}} \sum_{i=1}^{N_a} Z_q(\Psi^i) \ldots \sum_{j=1}^{N_a} Z_q(\Psi^j) \sum_{m=1}^{N_a}  Z_q(\Psi^k) \label{equ:proof1c}\\
  & =  2^{(d-1)(q-1)} \langle Z_q^{1\mathrm{d}} \rangle ^d \sim  N_a^{-d\tau_q^{1\mathrm{d}}} \sim \left( V_a^{d\mathrm{d}} \right)^{-\tau_q^{1\mathrm{d}}} \label{equ:proof1f}\;.
 \end{align}
\end{subequations}}
However, the separation in $2^{d-1}$ identical lattices used here cannot be exactly applied to the other sequences. In the following we will derive the equations for other metallic-mean models.

\subsection{General case -- arbitrary metallic-mean chains}

For some of the metallic-mean models even as well as odd chain lengths $N$ occur, where the eigenstates show some significant
differences for these two models as outlined in Sec.\ \ref{sec:construction}. Thus, for the proof we distinguish between even and odd approximants.

\paragraph{Even chain lengths $N_a$:}
We start with the easier case of even chain lengths $N$. In this case we can use an analogous derivation as for the silver-mean model.
Arbitrary metallic-mean chains are no palindroms and, thus, the grid in $d$ dimensions does not resolve into identical labyrinth tilings. However, due to the hidden mirror symmetries of the chains as discussed in Sec.\ \ref{sec:labyrinth} the wave functions of the different labyrinth tilings approach each other for $N \gg 1$, because then the influence of the boundaries and the single phason flip for $b > 2$ becomes negligible. Hence, the step from Eq.\ \eqref{equ:participation.2b} to Eq.\ \eqref{equ:participation.2c} is still approximately correct for $N \gg 1$.

\begin{figure}[b!]
 \centering
    \begin{overpic}[width=0.92\columnwidth]{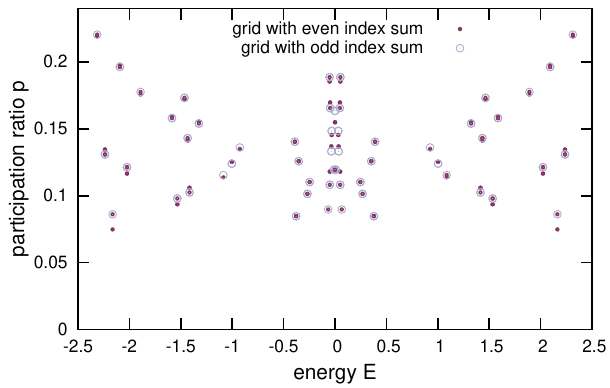} \put(2,3){\footnotesize{(a)}} \end{overpic}
    \begin{overpic}[width=0.92\columnwidth]{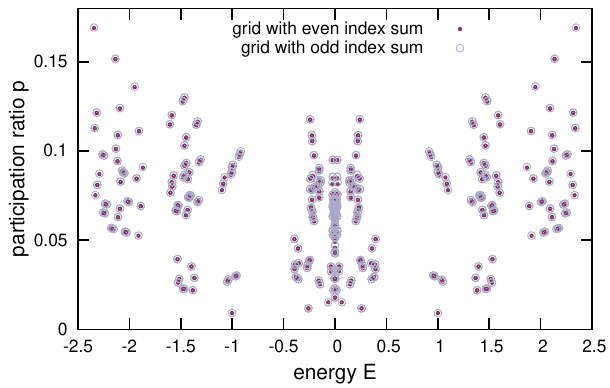} \put(2,3){\footnotesize{(b)}} \end{overpic}
    \begin{overpic}[width=0.92\columnwidth]{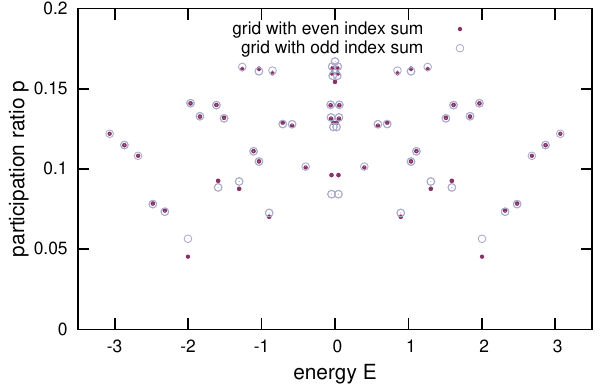} \put(2,3){\footnotesize{(c)}} \end{overpic}
    \begin{overpic}[width=0.92\columnwidth]{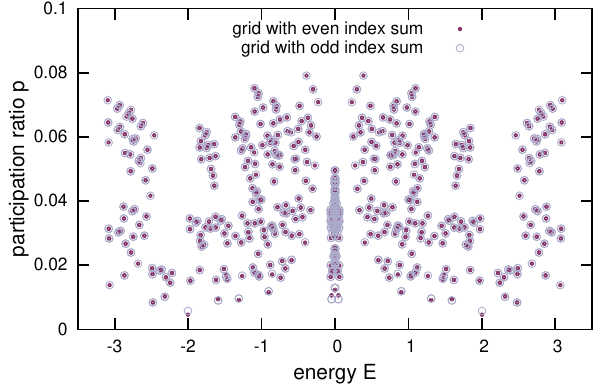} \put(2,3){\footnotesize{(d)}} \end{overpic}
  \caption{Difference of the participation ratios $p(\Phi^\mathbf{s})$ for the two possible labyrinth tilings, $\mathcal{L}$ and $\mathcal{L}^\star$, for the golden- and bronze-mean model in two dimensions for a coupling parameter $v = 0.3$.  Results are shown for $\mathcal{L}_6^{\mathrm{Au}}$ with $N = 14$ (a), $\mathcal{L}_8^{\mathrm{Au}}$ with $N = 35$ (b), $\mathcal{L}_3^{\mathrm{Br}}$ with $N = 14$ (c), and $\mathcal{L}_4^{\mathrm{Br}}$ with $N = 44$ (d).}
 \label{fig:deviation-p}
\end{figure}

The deviations of the average participation ratios for the two possible tilings in two dimensions, $\mathcal{L}$ and $\mathcal{L}^\star$, are shown for different system sizes in Fig.\ \ref{fig:deviation-p}. The results are quite close for both tilings of the golden- and bronze-mean model, respectively. Further, the deviations of the participation ratio for both tilings become smaller for higher approximants. To be specific, the total difference of the average participation ratio is smaller than 1\% for systems sizes $V_6^\mathrm{Au} = (N_6^\mathrm{Au})^2 / 2 = 98$ for the golden-mean model and $V_3^\mathrm{Bz} = 98$ for the bronze-mean model.

\paragraph{Odd chain lengths $N_a$:}

In the next step we have to derive such a relation also for odd system sizes. In this case the proof is more complicated because we have to treat the state $E^M = 0$ separately due to the special structure of the corresponding wave function (cp.\ Eq.\ \eqref{equ:octonacci.7}). In the following we show the derivation for the two-dimensional case, for which the grid decomposes into the two tilings $\mathcal{L}$ and $\mathcal{L}^\star$, where the state $E^M = 0$ belongs to the first one. This reasoning can be extended to arbitrary dimensions in a straightforward way.

Starting with the definition of the average generalized inverse participation number we obtain the scaling law from the following derivations:
\small
{\allowdisplaybreaks
\begin{subequations}
 \begin{align}
   &\langle Z_q^{2\mathrm{d}} \rangle = \frac{1}{V_a^{2\mathrm{d}}} \sum_{i,j \in \mathcal{L}} Z_q(\Phi^{ij}) \label{equ:proof2d-2a}\\
    &\stackrel{\eqref{equ:labyrinth.4b}}{=}  \frac{2}{N_a^2+1} \left[ \sum_{i=1}^{M-1} \sum_{j=1}^{N_a} Z_q(\Phi^{ij}) + \sum_{j=1}^{M} Z_q(\Phi^{Mj}) \right]\label{equ:proof2d-2b} \\
    &\stackrel{\eqref{equ:octonacci.6}}{=}  \frac{1}{N_a^2+1} \left[ \sum_{i=1}^{M-1} \sum_{j=1}^{N_a} Z_q(\Phi^{ij}) +  \sum_{i=M+1}^{N_a}\sum_{j=1}^{N_a} Z_q(\Phi^{ij}) \right]  + \nonumber \\
    & \hspace{0.7cm} \frac{1}{N_a^2+1} \left[\sum_{j=1}^{M-1} Z_q(\Phi^{Mj}) + \sum_{j=M+1}^{N_a} Z_q(\Phi^{Mj}) \right] + \nonumber \\
    & \hspace{0.7cm} \frac{2}{N_a^2+1} Z_q(\Phi^{MM})  \label{equ:proof2d-2c}\\
    &\stackrel{\eqref{equ:participation.1}}{=}  \frac{1}{N_a^2+1} \sum_{i=1 \atop i\ne M}^{N_a} \hspace{-0.15cm} \sum_{j=1 \atop \vee j\ne M}^{N_a} \sum_{l,m \in \mathcal{L}} \abs{\Phi_{lm}^{ij}}^{2q} + \frac{2}{N_a^2+1} \sum_{l,m \in \mathcal{L}} \abs{\Phi_{lm}^{MM}}^{2q} \label{equ:proof2d-2d}\\
    &\stackrel{N_a \gg 1}{\approx} \frac{1}{N_a^2+1} \sum_{i=1 \atop i\ne M}^{N_a} \hspace{-0.15cm} \sum_{j=1 \atop \vee j\ne M}^{N_a} \left[ \frac{1}{2}  \sum_{l,m \in \mathcal{L}} \abs{\Phi_{lm}^{ij}}^{2q} + \frac{1}{2} \sum_{l,m \in \mathcal{L}^\star} \abs{\Phi_{lm}^{ij}}^{2q} \right] + \nonumber \\
    & \hspace{0.7cm} \frac{2}{N_a^2+1} \Bigg[ \sum_{l,m \in \mathcal{L}} \abs{\Phi_{lm}^{MM}}^{2q}  + \underbrace{\sum_{l,m \in \mathcal{L}^\star} \abs{\Phi_{lm}^{MM}}^{2q}}_{=0} \Bigg] \label{equ:proof2d-2f}\\
    &\stackrel{\eqref{equ:labyrinth.3b}}{=} \frac{1}{2N_a^2+2} \sum_{i=1 \atop i\ne M}^{N_a} \hspace{-0.15cm} \sum_{j=1 \atop \vee j\ne M}^{N_a} \sum_{l=1}^{N_a} \sum_{m=1}^{N_a} \abs{\sqrt{2} \Psi_l^i \Psi_m^j}^{2q} + \nonumber \\
    & \hspace{0.7cm} \frac{2}{N_a^2+1} \sum_{l=1}^{N_a} \sum_{m=1}^{N_a} \abs{\Psi_l^M \Psi_m^M}^{2q} \label{equ:proof2d-2g}\\
    &= \frac{2^{q-1}}{N_a^2+1} \sum_{i=1}^{N_a} \sum_{j=1}^{N_a} \sum_{l=1}^{N_a} \sum_{m=1}^{N_a} \abs{\Psi_l^i \Psi_m^j}^{2q} +\nonumber \\
    & \hspace{0.7cm} \frac{2-2^{q-1}}{N_a^2+1} \sum_{l=1}^{N_a} \sum_{m=1}^{N_a} \abs{\Psi_l^M \Psi_m^M}^{2q} \label{equ:proof2d-2h}\\
    &= \frac{2^{q-1} N_a^2}{N_a^2+1} \frac{1}{N_a} \sum_{i=1}^{N_a} \sum_{l=1}^{N_a} \abs{\Psi_l^i}^{2q}  \frac{1}{N_a} \sum_{j=1}^{N_a} \sum_{m=1}^{N_a} \abs{\Psi_m^j}^{2q} + \nonumber\\
    & \hspace{0.7cm} \frac{2-2^{q-1}}{N_a^2+1} \sum_{l=1}^{N_a} \abs{\Psi_l^M}^{2q} \sum_{m=1}^{N_a} \abs{\Psi_m^M}^{2q} \label{equ:proof2d-2i}\\
    &\stackrel{N_a \gg 1}{\approx} 2^{q-1} \underbrace{\frac{N_a^2}{N_a^2+1}}_{\approx 1} \langle Z_q^{1\mathrm{d}} \rangle^2  +  \underbrace{\frac{2-2^{q-1}}{N_a^2+1}}_{\approx 0} Z_q(\Psi^M)^2 \label{equ:proof2d-2j}\\
    &\sim N_a^{-2\tau_q^{1\mathrm{d}}} \sim  \left(V_a^{2\mathrm{d}}\right)^{-\tau_q^{1\mathrm{d}}} \label{equ:proof2d-2k}\;.
 \end{align}
\end{subequations}}
\normalsize

The step from Eq.\ \eqref{equ:proof2d-2b} to Eq.\ \eqref{equ:proof2d-2c} is possible due to the same arguments as in the case with even chain lengths $N_a$. In Eq.\ \eqref{equ:proof2d-2g} we have to take into account that the state $i=j=M$ only differs from zero either on $\mathcal{L}$ or $\mathcal{L}^\star$. In our case this means that only the product of the $\Psi^{M+}$ state of Eq.\ \eqref{equ:octonacci.7} is considered for the state $E^{MM}$ because $\Psi^{M-}$ vanishes on the labyrinth tiling $\mathcal{L}$. All other steps of the proof follow the same way as for even system size $N_a$. Analogously the same concept can be applied to higher dimensions.

Hence, we have shown that the scaling of the average inverse participation numbers $\langle Z_q \rangle$ with the linear system size $N \gg 1$ is described by the one-dimensional scaling exponent according to
\begin{equation}
  \label{equ:relation}
  D_q^{d\mathrm{d}} = d D_q^\mathrm{1d} \;.
\end{equation}
Since the system size grows also with $V_{d\mathrm{d}} \sim N^d$, the eigenstates in these metallic-mean systems maintain their fractal nature, i.e.\ wave functions possess the same scaling exponent in arbitrary dimensions with respect to the actual system size $V_{d\mathrm{d}}$.

\section{Conclusion}

In this paper we have investigated the spatial distribution of the wave functions for the generalized labyrinth tilings in different dimensions $d$. First we have introduced the construction rules for the eigenstates in labyrinth tilings in two and three dimensions. The numerical results for the scaling behavior of the generalized inverse participation numbers $Z_q$ revealed the multifractal properties of the wave functions. In particular, we found numerical evidence that the spatial distribution of the wave functions, which is described by the scaling exponents $\gamma$ and $D_q$, is independent of the dimension $d$. This result has been substantiated by showing analytically that for the silver-mean model the scaling exponents $D_q$ in $d$ dimensions are always multiples of the one-dimensional scaling exponents $D_q^\mathrm{1d}$ according to Eq.\ \eqref{equ:relation}. For the other metallic-mean chains we have shown that the scaling exponents approach this relation asymptotically. This result shows that the wave functions in the metallic-mean systems are characterized by the same spatial expansion and the same multifractal properties in each dimension. Hence, the generalized labyrinth tilings with their special structures can be useful toy models to study the characteristics of higher-dimensional quasiperiodic systems efficiently by numerical methods and also analytically.

\bibliographystyle{unsrt}
\bibliography{literature}

\end{document}